\documentclass[runningheads]{llncs}

\usepackage[misc]{ifsym}
\usepackage[colorinlistoftodos]{todonotes}
\usepackage{amssymb}

\usepackage[caption=false]{subfig}
\usepackage{float}

\usepackage{multirow}
\usepackage{rotating}

\begin{document}

\title{GAN-Based Multiple Adjacent Brain MRI\\ Slice Reconstruction for Unsupervised \\Alzheimer's Disease Diagnosis}
\titlerunning{GAN-Based Unsupervised Alzheimer's Disease Diagnosis}  
%
\author{Changhee Han\inst{1,2,3}(\Letter) \and Leonardo Rundo\inst{4,5} \and Kohei Murao\inst{1} \and \\ Zolt\'{a}n \'{A}d\'{a}m Milacski\inst{6} \and Kazuki Umemoto\inst{7} \and Evis Sala\inst{4,5} \and \\ Hideki Nakayama\inst{3,8} \and Shin'ichi Satoh\inst{1}}
\authorrunning{C. Han et al.} 
%
%
\institute{Research Center for Medical Big Data,\\National Institute of Informatics, Tokyo, Japan\\
\email{han@nlab.ci.i.u-tokyo.ac.jp}, \and
LPixel Inc., Tokyo, Japan
\and
Graduate School of Information Science and Technology,\\The University of Tokyo, Tokyo, Japan \and
Department of Radiology, University of Cambridge, Cambridge, United Kingdom \and
Cancer Research UK Cambridge Centre, Cambridge, United Kingdom \and Department of Artificial Intelligence,\\ELTE E\"{o}tv\"{o}s Lor\'{a}nd University, Budapest, Hungary \and Department of Rehabilitation Medicine,\\Juntendo University School of Medicine, Tokyo, Japan \and International Research Center for Neurointelligence (WPI-IRCN), The University of Tokyo Institutes for Advanced Study, The University of Tokyo, Tokyo, Japan}

\maketitle              

\begin{abstract}
Unsupervised learning can discover various unseen diseases, relying on large-scale unannotated medical images of healthy subjects. Towards this, unsupervised methods reconstruct a single medical image to detect outliers either in the learned feature space or from high reconstruction loss. However, without considering continuity between multiple adjacent slices, they cannot directly discriminate diseases composed of the accumulation of subtle anatomical anomalies, such as Alzheimer's Disease (AD). Moreover, no study has shown how unsupervised anomaly detection is associated with disease stages. Therefore, we propose a two-step method using Generative Adversarial Network-based multiple adjacent brain MRI slice reconstruction to detect AD at various stages: (\textit{Reconstruction}) Wasserstein loss with Gradient Penalty $+$ $\ell _1$ loss---trained on $3$ healthy slices to reconstruct the next $3$ ones---reconstructs unseen healthy/AD cases; (\textit{Diagnosis}) Average/Maximum loss (e.g., $\ell _2$ loss) per scan discriminates them, comparing the reconstructed/ground truth images. The results show that we can reliably detect AD at a very early stage with Area Under the Curve (AUC) $0.780$ while also detecting AD at a late stage much more accurately with AUC $0.917$; since our method is fully unsupervised, it should also discover and alert any anomalies including rare disease.

\end{abstract}
\section{Introduction}
\label{sec:Intro}

Deep Learning can achieve accurate computer-assisted diagnosis when large-scale annotated training samples are available. In medical imaging, unfortunately, preparing such massive annotated datasets is often unfeasible; to tackle this important problem, researchers have proposed various data augmentation techniques, including Generative Adversarial Network (GAN)-based ones~\cite{goodfellow2014,FridAdar,Han2,Han3,han2019CIKM}. However, even exploiting these techniques, supervised learning still requires many images with pathological features, even for rare disease, to make a reliable diagnosis; nevertheless, it can only detect already-learned specific pathologies. In this regard, as physicians notice previously unseen anomaly examples using prior information on healthy body structure, unsupervised anomaly detection methods leveraging only large-scale healthy images can discover and alert unseen disease when their generalization fails.

\begin{figure}[t]
  \centering
  \centerline{\includegraphics[width=1\columnwidth]{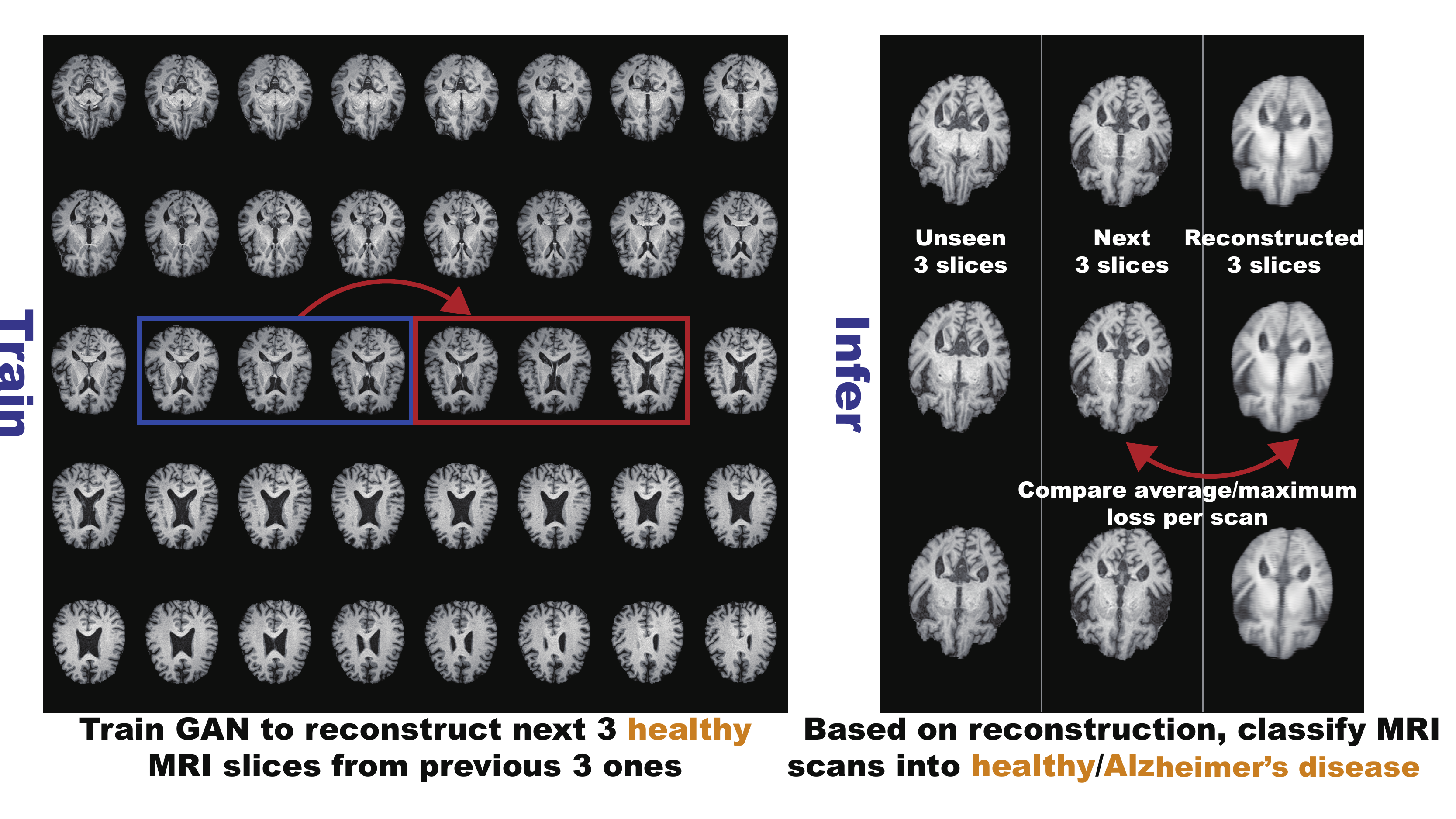}}
\caption{Unsupervised AD diagnosis framework: we train WGAN-GP $+$ $\ell _1$ loss on $3$ healthy brain axial MRI slices to reconstruct the next $3$ ones, and test it on both unseen healthy and AD cases to classify them based on average/maximum loss (e.g., $\ell _2$ loss) per scan.}
\label{fig1}
\vspace{-0.3cm}
\end{figure}

Towards this, researchers reconstructed a single medical image \textit{via} GANs~\cite{Schlegl}, AutoEncoders (AEs)~\cite{Uzunova}, or combining them, since GANs can generate realistic images and AEs, especially Variational AEs, can directly map data onto its latent representation~\cite{Chen}; then, unseen images were scored by comparing them with reconstructed ones to discriminate a pathological image distribution (i.e., outliers either in the learned feature space or from high reconstruction loss). However, those single image reconstruction methods mainly target diseases easy-to-detect from a single image even for non-expert human observers, such as glioblastoma on Magnetic Resonance (MR) images~\cite{Chen} and lung cancer on Computed Tomography images~\cite{Uzunova}. Without considering continuity between multiple adjacent images, they cannot directly discriminate diseases composed of the accumulation of subtle anatomical anomalies, such as Alzheimer's Disease (AD). Moreover, no study has shown so far how unsupervised anomaly detection is associated with disease stages.
We thus propose a two-step method using GAN-based multiple adjacent brain MRI slice reconstruction to detect AD at various stages (Fig.~\ref{fig1}): (\textit{Reconstruction}) Wasserstein loss with Gradient Penalty (WGAN-GP)~\cite{Gulrajani,han2018} $+$ $\ell _1$ loss---trained on $3$ healthy brain axial MRI slices to reconstruct the next $3$ ones---reconstructs unseen healthy/AD cases; (\textit{Diagnosis}) Average/Maximum loss (e.g., $\ell _2$ loss) per scan discriminates them, comparing the reconstructed and ground truth images.\\

\paragraph{Contributions.} Our main contributions are as follows:
\begin{itemize}
\item \textbf{MRI Slice Reconstruction:} This first multiple MRI slice reconstruction approach can predict the next $3$ brain MRI slices from the previous $3$ ones only for unseen images similar to training data by combining WGAN-GP and $\ell _1$ loss.

\item \textbf{Unsupervised Anomaly Detection:} This first unsupervised anomaly detection across different disease stages reveals that, like physicians' way of diagnosis, massive healthy data can reliably aid early diagnosis, such as of MCI, while also detecting late-stage disease much more accurately by discriminating with $\ell _2$ loss.

\item \textbf{Alzheimer's Disease Diagnosis:} This first unsupervised AD diagnosis study can reliably detect AD and also any other diseases.

\end{itemize}

The remainder of the manuscript is organized as follows: Sect.~\ref{sec:Background} outlines the state-of-the-art of automated AD diagnosis; Sect.~\ref{sec:MatMet} describes the analyzed MRI dataset, as well as the proposed GAN-based unsupervised AD diagnosis framework; experimental results are shown and discussed in Sect. \ref{sec:Results}; finally, Sect.~\ref{sec:Conclusion} provides conclusive remarks and future work.

\section{Automated Alzheimer's Disease Diagnosis}
\label{sec:Background}

Despite the clinical, social, and economic significance of early AD diagnosis---primarily associated with MCI detection---it generally relies on subjective assessment by physicians (e.g., neurologists, geriatricians, and psychiatrists); to tackle this open challenge, researchers have used classic supervised Machine Learning based on hand-crafted features~\cite{salvatore2015,nanni2019}. More recently, Deep Learning has attracted great attentions owing to its more abstract and descriptive embedding based on multiple non-linear transformations: Liu \textit{et al.} used a semi-supervised CNN to significantly reduce the need for labeled training data\cite{liu2014early}; for clinical decision-making, Suk \textit{et al.} integrated multiple sparse regression models (namely, Deep Ensemble Sparse Regression Network)~\cite{suk2017}; Spasov \textit{et al.} devised a parameter-efficient CNN for 3D separable convolutions, combining dual learning and a specific layer to predict the conversion from MCI to AD within $3$ years~\cite{spasov2019}; instead of exploiting the CNNs, Parisot used a semi-supervised Graph Convolutional Network trained on a sub-set of labeled nodes with diagnostic outcomes to represent sparse clinical data~\cite{parisot2018}.

However, to the best of our knowledge, no existing work has conducted fully unsupervised anomaly detection for AD diagnosis since capturing subtle anatomical differences between MCI and AD is challenging.

\section{Materials and Methods}
\label{sec:MatMet}

\subsection{OASIS-3 Dataset}
\label{sec:dataset}

We use a longitudinal dataset of $176 \times 240$/$176 \times 256$ T1-weighted (T1w) 3T brain axial MRI slices containing both normal aging subjects/AD patients extracted from the Open Access Series of Imaging Studies-3 (OASIS-3)~\cite{lamontagne2018oasis}. The $176 \times 240$ slices are zero-padded to reach $176 \times 256$ pixels. Relying on Clinical Dementia Rating (CDR)~\cite{Morris}, common clinical scale for the staging of dementia, the subjects are comprised of:

\begin{itemize}
\item Unchanged CDR $= 0$: Cognitively healthy population;
\item CDR $= 0.5$: Very mild dementia\ ($\sim$ MCI);
\item CDR $= 1$: Mild dementia;
\item CDR $= 2$: Moderate dementia.
\end{itemize}

Since our dataset is longitudinal and the same subject's CDRs may vary (e.g., CDR $= 0$ to CDR $= 0.5$), we only use scans with unchanged CDR $= 0$ to assure certainly healthy scans. As CDRs and MRI scans are not always simultaneously acquired, we label MRI scans with CDRs at the closest date. We only select brain MRI slices including hippocampus/amygdala/ventricles among whole $256$ axial slices per scan to avoid over-fitting from AD-irrelevant information; the atrophy of the hippocampus/amygdala/cerebral cortex, and enlarged ventricles are strongly associated with AD, and thus they mainly affect the AD classification performance of Machine Learning~\cite{Ledig}. Moreover, we discard low-quality MRI slices. The remaining dataset is divided as follows:

\begin{itemize}
\item Training set: Unchanged CDR $= 0$ ($408$ subjects/$1,133$ scans/$57,834$ slices);
\item Validation set: Unchanged CDR $= 0$ ($55$ subjects/$155$ scans/$8,080$ slices), \\CDR $= 0.5$ ($53$ subjects/$85$ scans/$4,607$ slices),\\CDR $= 1$ ($29$ subjects/$45$ scans/$2,518$ slices),\\CDR $= 2$ ($2$ subjects/$4$ scans/$160$ slices);
\item Test set: Unchanged CDR $= 0$ ($113$ subjects/$318$ scans/$16,198$ slices),\\CDR $= 0.5$ ($99$ subjects/$168$ scans/$9,206$ slices),\\CDR $= 1$ ($61$ subjects/$90$ scans/$5,014$ slices),\\CDR $= 2$ ($4$ subjects/$6$ scans/$340$ slices).
\end{itemize}
The same subject's scans are included in the same dataset. The datasets are strongly biased towards healthy scans similarly to MRI inspection in the clinical routine. During training for reconstruction, we only use the training set containing healthy slices to conduct unsupervised learning.

\subsection{GAN-based Multiple Adjacent Brain MRI Slice Reconstruction}
\label{sec:MRIrecon}

To model strong consistency in healthy brain anatomy (Fig.~\ref{fig1}), in each scan, we reconstruct the next $3$ MRI slices from the previous $3$ ones using an image-to-image GAN (e.g., if a scan includes $40$ slices $s_i$ for $i=1,\dots,40$, we reconstruct all possible $35$ setups: $(s_i)_{i\in\{1,2,3\}} \mapsto (s_i)_{i\in\{4,5,6\}}$; $(s_i)_{i\in\{2,3,4\}} \mapsto (s_i)_{i\in\{5,6,7\}}$; \dots; $(s_i)_{i\in\{35,36,37\}} \mapsto (s_i)_{i\in\{38,39,40\}}$). We concatenate adjacent $3$ grayscale slices into $3$ channels, such as in RGB images. The GAN uses a U-Net-like~\cite{Ronneberger,RundoUSEnet} generator with $4$ convolutional layers in encoders and $4$ deconvolutional layers in decoders respectively with skip connections as well as a discriminator with $3$ decoders. We apply batch normalization to both convolution with Leaky Rectified Linear Unit (ReLU) and deconvolution with ReLU. To confirm how reconstructed images' realism and anatomical continuity affect anomaly detection, we compare the GAN models with different loss functions: (\textit{i}) Dice loss (i.e., a plain U-Net without the discriminator); (\textit{ii}) WGAN-GP loss; (\textit{iii}) WGAN-GP loss $+$ 100 $\ell _1$ loss.
Among 8 losses comparing ground truth/reconstructon, average $\ell _2$ loss per scan always outperforms the other losses during validation for U-Net and WGAN-GP without/with $\ell _1$ loss, and thus we use this loss for testing.

\paragraph{Implementation Details} Considering its computational speed, U-Net training lasts for $600,000$ steps with a batch size of $64$ and both GAN trainings last for $300,000$ steps with a batch size of $32$. We use $2.0 \times 10^{-4}$ learning rate for the Adam optimizer~\cite{kingma2014}. The framework is implemented on Keras with TensorFlow as backend.

\subsection{Unsupervised Alzheimer's Disease Diagnosis}
\label{sec:ADdiagnosis}

During validation, we compare the following average/maximum losses per scan (i.e., $8$ losses) between reconstructed/ground truth 3 slices (Fig.~\ref{fig1}): (\textit{i}) $\ell _1$ loss; (\textit{ii}) $\ell _2$ loss; (\textit{iii}) Dice loss; (\textit{iv}) Structural Similarity loss. For each model's testing, we separately pick the loss showing the highest AUC between CDR = 0 (i.e., healthy population) \textit{vs} all the other CDRs (i.e., dementia) during validation. As a result, we pick the average $\ell _2$ loss per scan for all models since squared error is sensitive to outliers and it always outperforms the others. To evaluate its unsupervised AD diagnosis performance for test sets, we show Receiver Operating Characteristics (ROCs)/AUCs between CDR $= 0$ \textit{vs} (\textit{i}) all the other CDRs; (\textit{ii}) CDR $= 0.5$; (\textit{iii}) CDR $= 1$; (\textit{iv}) CDR $= 2$. We visualize $\ell _2$ loss distributions of CDR $= 0/0.5/1/2$ to know how disease stages affect its discrimination.

\begin{figure}[t]
  \centering
  \centerline{\includegraphics[width=\columnwidth]{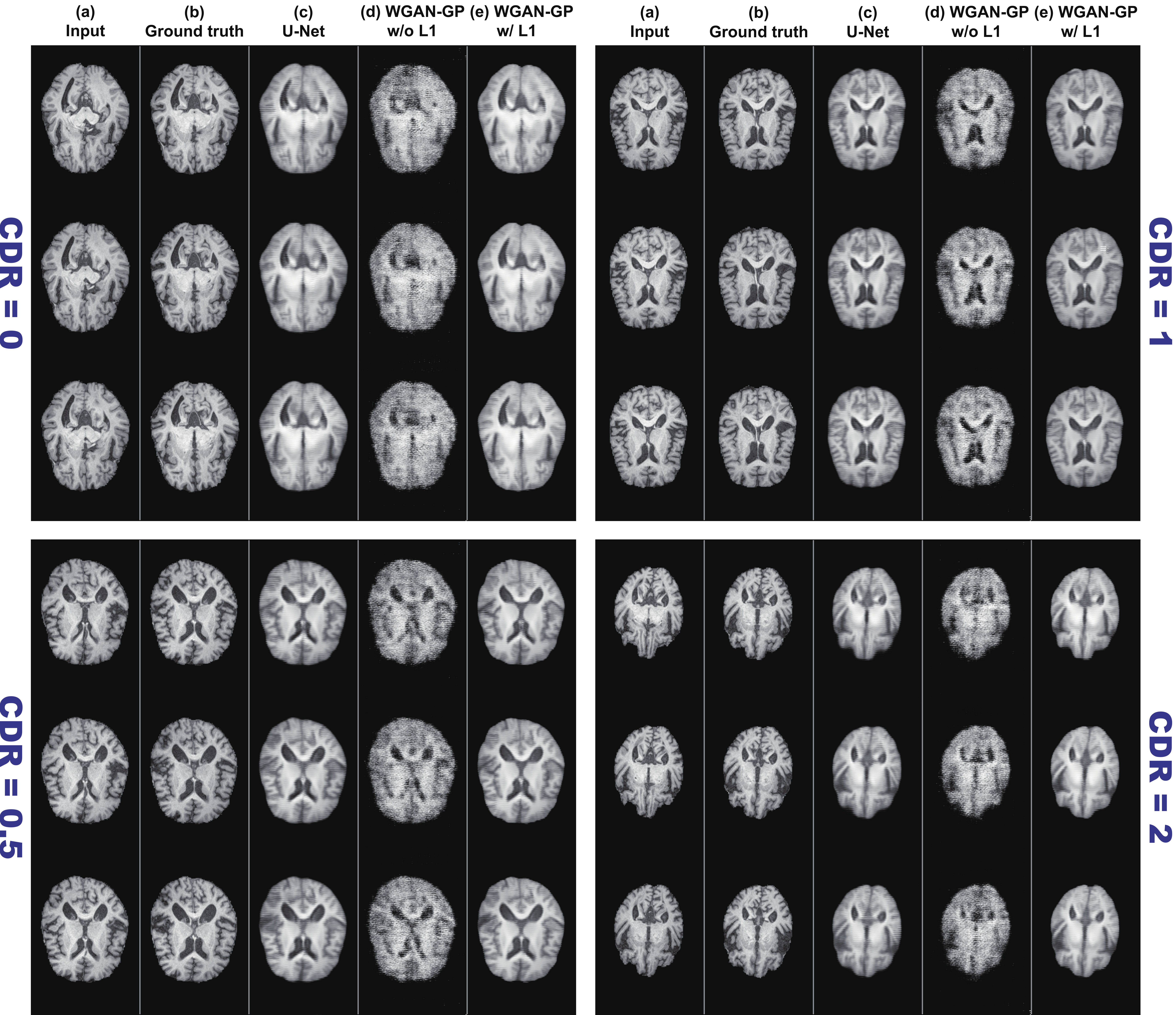}}
\caption{Example brain MRI slices with CDR $= 0/0.5/1/2$ from test sets: (a) Input $3$ real slices; (b) Ground truth next $3$ real slices; (c) Next $3$ slices reconstructed by U-Net; (d), (e) Next $3$ slices reconstructed by WGAN-GP without/with $\ell _1$ loss.}
\label{fig2}
\end{figure}

\begin{figure}[t]
  \centering
  \centerline{\includegraphics[width=0.95\columnwidth]{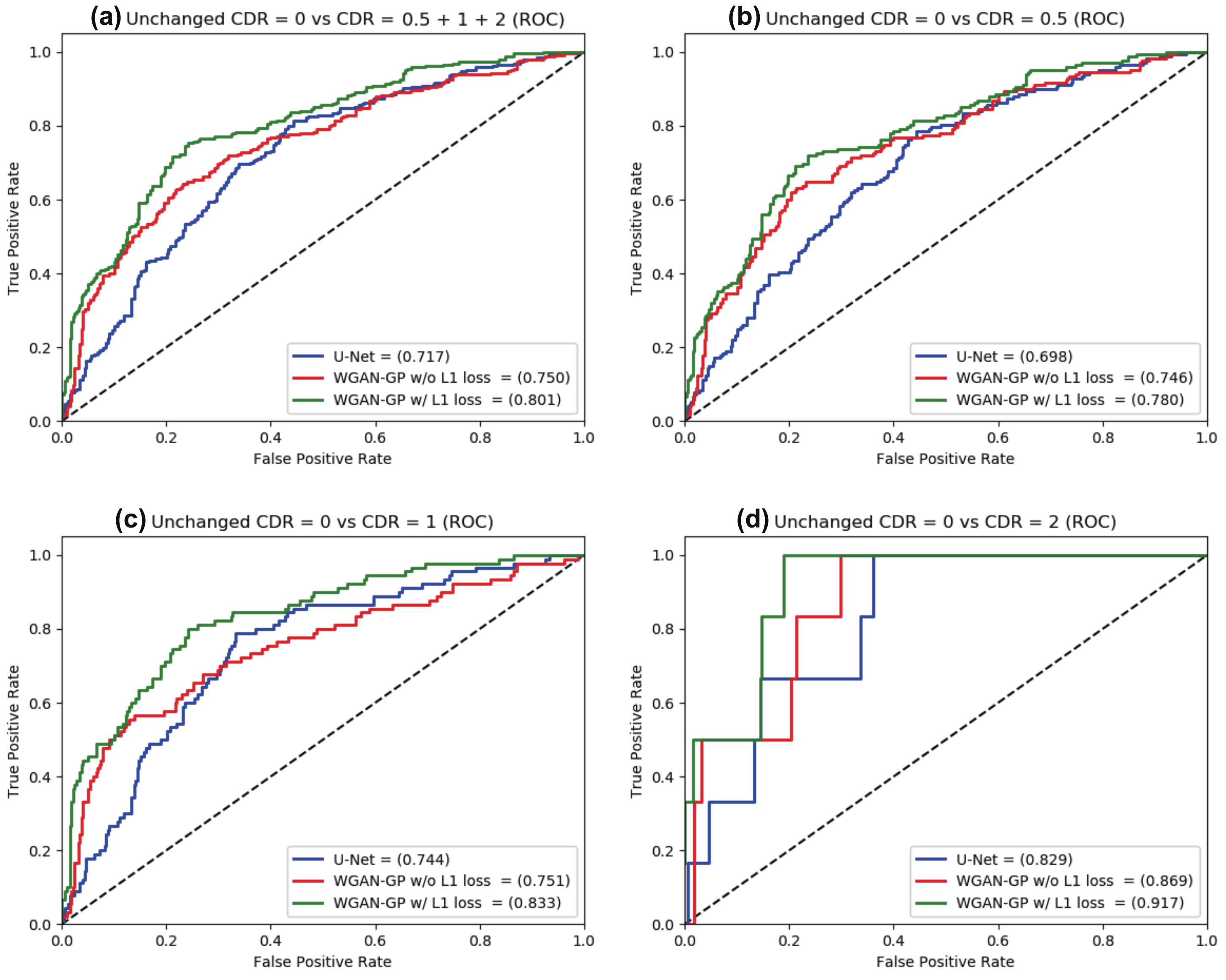}}
\caption{Unsupervised anomaly detection results using average $\ell _2$ loss per scan on reconstructed brain MRI slices (ROCs and AUCs): unchanged CDR $= 0$ (i.e., cognitively healthy population) is compared with: (a) all the other CDRs (i.e., dementia); (b) CDR $= 0.5$ (i.e., very mild dementia); (c) CDR $= 1$ (i.e., mild dementia); (d) CDR $= 2$ (i.e., moderate dementia).}
\label{fig3}
\end{figure}

\begin{figure}[t!]
  \centering
  \centerline{\includegraphics[width=0.62\columnwidth]{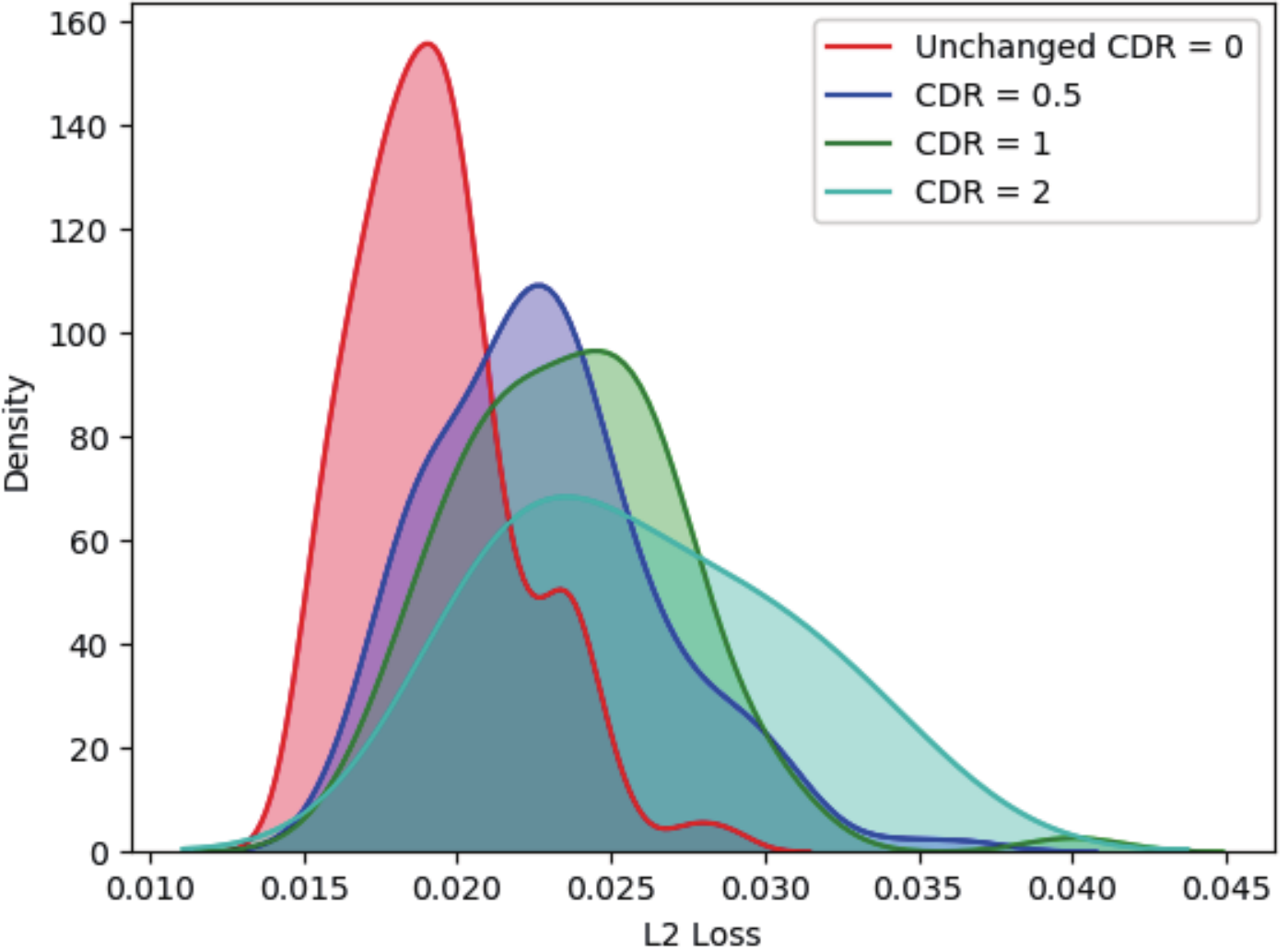}}
\caption{Distributions of average $\ell _2$ loss per scan evaluated on brain MRI slices with CDR $= 0/0.5/1/2$ reconstructed by WGAN-GP $+$ $\ell _1$ loss.}
\label{fig4}
\end{figure}

\section{Results}
\label{sec:Results}

\subsection{\textbf{Reconstructed Brain MRI Slices}}
\label{sec:MRIreconRes}

Fig.~\ref{fig2} illustrates example real MRI slices from test sets and their reconstruction by U-Net and WGAN-GP without/with $\ell _1$ loss. The WGAN-GP $+$ $\ell _1$ loss can successfully capture T1w-specific appearance and anatomical changes from the previous $3$ slices more smoothly than the U-Net and in more detail than the WGAN-GP without $\ell _1$ loss. Since the models are trained only on healthy slices, reconstructing slices with higher CDRs tends to comparatively fail, especially around hippocampus, amygdala, cerebral cortex, and ventricles due to their insufficient atrophy after reconstruction.

\subsection{Unsupervised AD Diagnosis Results}
\label{sec:ADdiagnosisRes}

Fig.~\ref{fig3} shows ROC curves and their AUCs of unsupervised anomaly detection. Since brains with higher CDRs accompany stronger anatomical atrophy from healthy brains, their AUCs between unchanged CDR = 0 remarkably increase as CDRs increase. Clearly outperforming the other methods in every condition, WGAN-GP $+$ $\ell _1$ loss achieves excellent AUCs, especially for higher CDRs---it obtains AUC $= 0.780/0.833/0.917$ for CDR $= 0$ \textit{vs} CDR $= 0.5/1/2$, respectively; this experimental finding derives from $\ell _1$ loss' good realism sacrificing diversity (i.e., generalizing well only for unseen images with a similar distribution to training images) and WGAN-GP loss' ability to capture recognizable structure. Fig.~\ref{fig4} indicates its good discrimination ability even between healthy subjects \textit{vs} MCI patients (i.e., CDR $= 0$ \textit{vs} CDR $= 0.5$), which is extremely difficult even in a supervised manner~\cite{Ledig}. Interestingly, unlike our visual expectation, WGAN-GP without $\ell _1$ loss outperforms plain U-Net regardless of its very blurred reconstruction, showing the superiority of GAN-based reconstruction for diagnosis.

\section{Conclusions and Future Work}
\label{sec:Conclusion}

Using a massive amount of healthy images, our GAN-based multiple MRI slice reconstruction can successfully discriminate AD patients from healthy subjects for the first time in an unsupervised manner; our solution leverages a two-step approach: (\textit{Reconstruction}) $\ell _1$ loss generalizes well only for unseen images with a similar distribution to training images while WGAN-GP loss captures recognizable structure; (\textit{Diagnosis}) $\ell _2$ loss clearly discriminates healthy/abnormal data as squared error becomes huge for outliers.
Using $1,133$ healthy MRI scans for training, our approach can reliably detect AD at a very early stage, Mild Cognitive Impairment (MCI), with Area Under the Curve (AUC) $0.780$ while detecting AD at a late stage much more accurately with AUC $0.917$---implying its ability to also detect any other diseases.

Accordingly, this first unsupervised anomaly detection across different disease stages reveals that, like physicians' way of diagnosis, large-scale healthy data can reliably aid early diagnosis, such as of MCI, while also detecting late-stage disease much more accurately. Since our method well detects the unseen disease hard-to-detect even in supervised learning, this should also discover/alert any anomalies including rare disease, where supervised learning is inapplicable. As future work, we will reconstruct slices from both previous/next $3$ slices (e.g.,  slices $s_i$ for $i=1,\dots,9$, $(s_i)_{i\in\{1,2,3,7,8,9\}} \mapsto (s_i)_{i\in\{4,5,6\}}$) for robustness, also optimizing the number of slices (e.g., 3 slices to 1 or 5 slices). We will investigate more reconstruction networks (e.g., GANs with attention mechanisms) and multiple loss functions for both reconstruction/diagnosis.
Lastly, we plan to detect and locate various diseases, including cancer \cite{rundo2018NC} and rare diseases---this work only uses brain MRI slices including hippocampus/amygdala/ventricles for AD diagnosis, but we may have to use all or most brain MRI slices to also detect anomalies appearing in other anatomical locations within the brain. Integrating multimodal imaging data, such as Positron Emission Tomography with specific radiotracers~\cite{rundoCMPB2017}, might further improve AD diagnosis~\cite{brier2016}, even when analyzed modalities are partially unavailable~\cite{li2014multimodal}.

\section*{Acknowledgment}
This research was supported by AMED Grant Number JP18lk1010028, and also partially supported by The Mark Foundation for Cancer Research and Cancer Research UK Cambridge Centre [C9685/A25177].
Additional support has been provided by the National Institute of Health Research (NIHR) Cambridge Biomedical Research Centre.
Zolt\'{a}n \'{A}d\'{a}m Milacski was supported by Grant Number VEKOP-2.2.1-16-2017-00006. The OASIS-3 dataset has Grant Numbers P50 AG05681, P01 AG03991, R01 AG021910, P50 MH071616, U24 RR021382, and R01 MH56584.

%
%

\bibliographystyle{splncs} 
\bibliography{biblio}

\end{document}